\documentclass{sig-alternate-05-2015}
\usepackage{hyperref}
\usepackage{multirow}

\begin{document}

% Copyright
%\setcopyright{acmcopyright}
%\setcopyright{acmlicensed}
\setcopyright{rightsretained}
%\setcopyright{usgov}
%\setcopyright{usgovmixed}
%\setcopyright{cagov}
%\setcopyright{cagovmixed}

% DOI
\doi{10.475/123_4}

% ISBN
\isbn{123-4567-24-567/08/06}
%Conference
\conferenceinfo{PLDI '13}{June 16--19, 2013, Seattle, WA, USA}

\acmPrice{\$15.00}

%
% --- Author Metadata here ---
\conferenceinfo{Under Review for CIKM}{2016, Indiana USA}

%\CopyrightYear{2007} % Allows default copyright year (20XX) to be over-ridden - IF NEED BE.
%\crdata{0-12345-67-8/90/01}  % Allows default copyright data (0-89791-88-6/97/05) to be over-ridden - IF NEED BE.
% --- End of Author Metadata ---

\title{Will Sanders Supporters Jump Ship for Trump? Fine-grained Analysis of Twitter Followers}

%\title{Will Sanders Supporters Jump Ship for Trump?\\ An Analysis of Sanders Followers on Twitter}


\numberofauthors{3} %  in this sample file, there are a *total*
\author{
% 1st. author
\alignauthor
{Yu Wang}\\
	   \affaddr{Political Science}\\
       \affaddr{University of Rochester}\\
       \affaddr{Rochester, NY, 14627}\\
       \email{yu.wang@rochester.edu}
         \alignauthor
      Yang Feng\\
       \affaddr{Computer Science}\\
       \affaddr{University of Rochester}\\
       \affaddr{Rochester, NY, 14627}\\
       \email{yfeng23@cs.rochester.edu}
                \alignauthor        
       Xiyang Zhang\\
	   \affaddr{Psychology}\\
       \affaddr{Beijing Normal University}\\
       \affaddr{Beijing, 100875}\\
       \email{zxy2013@mail.bnu.edu.cn}
       \and  
       \alignauthor   
       Richard Niemi\\
	   \affaddr{Political Science}\\
       \affaddr{University of Rochester}\\
       \affaddr{Rochester, NY, 14627}\\
       \email{niemi@rochester.edu}      
         \alignauthor
         Jiebo Luo\\
       \affaddr{Computer Science}\\
       \affaddr{University of Rochester}\\
       \affaddr{Rochester, NY, 14627}\\
       \email{jluo@cs.rochester.edu}        
     %  \alignauthor
       \iffalse
\alignauthor
         {*}\\
       \affaddr{*}\\
       \affaddr{*}\\
       \email{*}  
            \alignauthor   
         {*}\\
       \affaddr{*}\\
       \affaddr{*}\\
       \email{*}  
       \alignauthor   
         {*}\\
       \affaddr{*}\\
       \affaddr{*}\\
       \email{*}  
       \
       \and  
    \alignauthor   
         {*}\\
       \affaddr{*}\\
       \affaddr{*}\\
       \email{*}  
       \alignauthor   
         {*}\\
       \affaddr{*}\\
       \affaddr{*}\\
       \email{*}\\
       \fi
}
\maketitle
\begin{abstract}
In this paper, we study the likelihood of Bernie Sanders supporters voting for Donald Trump instead of Hillary Clinton. Building from a unique time-series dataset of the three candidates' Twitter followers, which we make public here, we first study the proportion of Sanders followers  who simultaneously follow Trump (but not Clinton) and how this evolves over time. Then we train a convolutional neural network to classify the gender of Sanders followers, and study whether men are more likely to jump ship for Trump than women. Our study shows that between March and May an increasing proportion of Sanders followers are following Trump (but not Clinton). The proportion of Sanders followers who follow Clinton but not Trump has actually decreased. Equally important, our study suggests that the jumping ship behavior will be affected by  gender and that men are more likely to switch to Trump than women.

\end{abstract}


%\section*{Categories and Subject Descriptors}
%H.3.5 [Information Systems]: Information Storage and Retrieval - Online Information Services.



\section{Introduction}

Even as Hillary Clinton moves closer and closer to clinch the Democratic nomination, some Sanders supporters insist that they will not vote for her in the general election. This leaves a golden opportunity to Donald Trump, the presumptive Republican nominee who has been sharing similar campaign messages with Sanders on trade and campaign finance. Trump, on the other hand, also makes it clear that he will target Sanders supporters.\footnote{Newsweek, http://www.newsweek.com/can-trump-win-over-bernie-sanders-supporters-459218.}

\iffalse Shortly after the Indiana Primary on May 3rd, Ted Cruz and John Kasich dropped out of the race, and Donald Trump emerged the presumptive Republican nominee for president. On the Democratic side, Hillary Clinton is also moving closer and closer to clinch the nomination. Trump's message is clear: he is going to win over Bernie supporters.\footnote{Newsweek: http://www.newsweek.com/can-trump-win-over-bernie-sanders-supporters-459218.} Trump's strategist Lewandowski suggests that Trump and Sanders share similar campaign messages and believes that Trump can win over Sanders supporters. \fi

%And a number of recent polls

This new dynamic quickly became hotly debated and the looming question is ``Can Trump win over Sanders supporters?''\footnote{New York Times, http://www.nytimes.com/2016/04/29/

us/politics/hillary-clinton-donald-trump-women.html.} A number of recent polls, including ABC/Washington Post, CBS/NYT and YouGov, do suggest that some Sanders supporters could end up voting for Trump and that this is particularly so for his male supporters.\footnote{New York Times, http://www.nytimes.com/2016/05/25/

upshot/explaining-hillary-clintons-lost-ground-in-the-polls.

html.} In this paper, we investigate this dynamic in Twitter. 

\begin{figure}[!h]
\centering
\includegraphics[height=5.1cm,width=8.4cm]{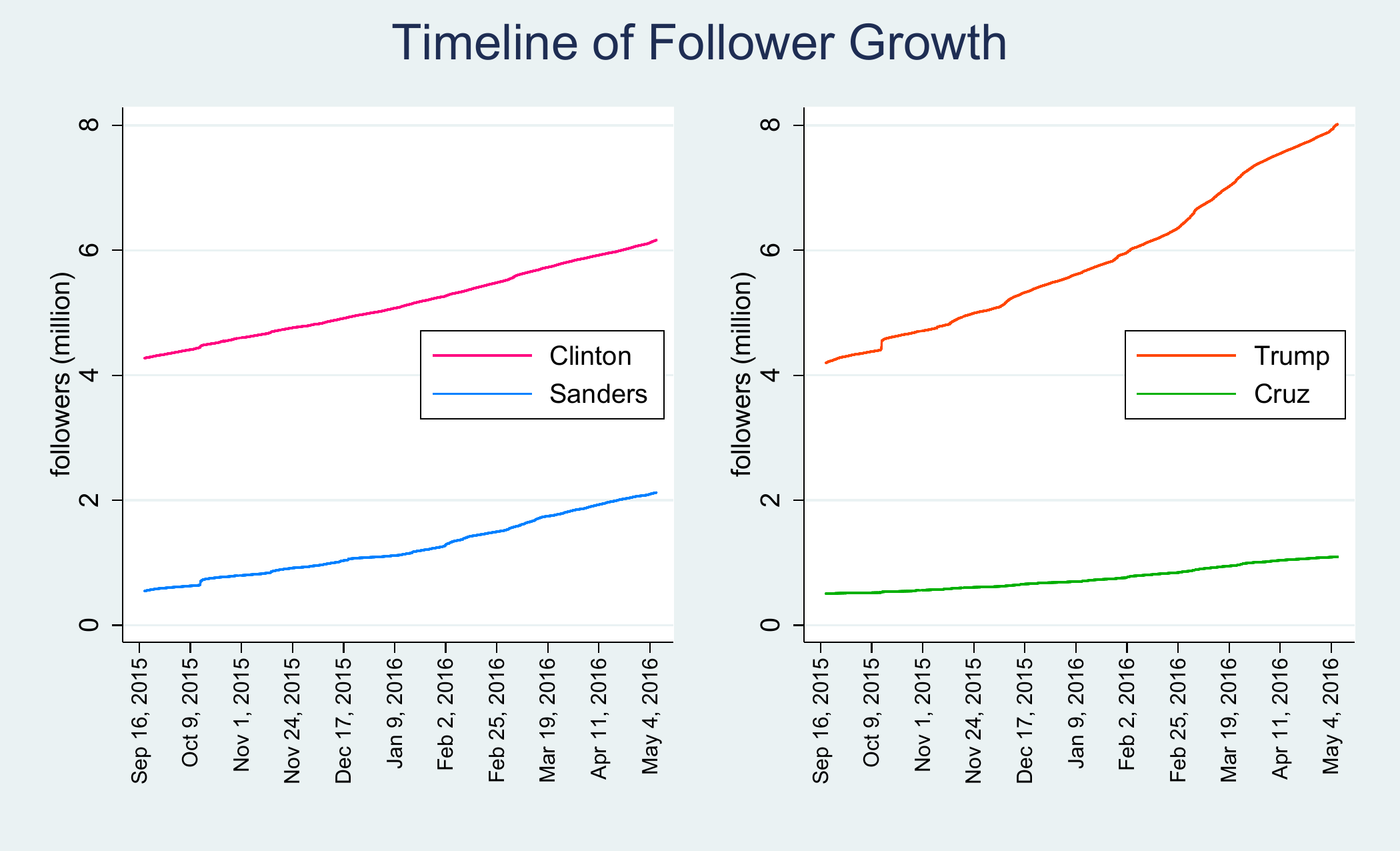}
\caption{Number of Followers for Hillary Clinton, Bernie Sanders, Donald Trump and Ted Cruz.}
\label{timeline}
\end{figure}

Building from a unique dataset of the three candidates' Twitter followers (Figure \ref{timeline}), we first examine the proportion of Sanders followers who simultaneously follow Trump or Clinton. Then we train a convolutional neural network to classify the followers' gender and study how men in particular are responding to Trump's overture. Our study suggests that in Twitter there also exists a shift towards Trump among Sanders followers, and that male Sanders followers are responding more positively to Trump than females.

\section{Related Literature}
Our work builds on previous literature in electoral studies, data mining, and computer vision.

In eletoral studies, researchers have argued that gender constitutes an important factor in voting behavior. One common observation is that women tend to vote for women, which is usually referred to as gender affinity effect \cite{sexAndGOP,genderAffinityEffect}. In the 2016 presidential election, Hillary Clinton also portrays herself as a champion ``fighting for women's healthcare and paid family leave and equal pay.'' Our work will test the strength of this gender affinity effect when Sanders supporters decide whether to jump ship for Trump or not.

In data mining, there is a burgeoning literature on using social media data to analyze and predict elections. In particular, several studies have explored ways to infer users' preferences. According to \cite{tweets2polls}, tweets with sentiment can potentially serve as votes and substitute traditional polling. \cite{trumponfire} exploits the variations in the number of `likes' of the tweets to infer Trump followers' topic preferences. \cite{facebookCongress} uses candidates' `likes' in Facebook to quantify a campaign's success in engaging the public. \cite{neco} uses follower growth on public debate dates to measure candidates' debate performance. Our work also pays close attention to the number of followers, but we go further by investigating the composition of these followers.

Our work also ties in with current computer vision research. In this dimension, our work is related to gender classification using facial features. \cite{israel} uses a five-layer network to classify both age and gender. \cite{ginosar} introduces a dataset of frontal-facing American high school yearbook photos and uses the extracted facial features to study historical trends in the U.S. \cite{facerace} provides a comprehensive survey of race classification based on facial features. \cite{trumpists} uses user profile images to study and compare the social demographics of Trump followers and Clinton followers. \cite{votingfeet} focuses specifically on the unfollowers of Donald Trump and Hillary Clinton and reports that women are more likely to unfollow both candidates. \cite{womancard} studies the `woman card' exchange between Trump and Clinton, and finds that the `woman card' exchange has made women more likely to follow Clinton and less likely to unfollow her.

\section{Data and Methodology}
In this section, we describe our dataset \textit{US2016}, the pre-processing procedures and our CNN model. One key variable is \textit{number of followers.} This variable is available for all three candidates and covers the entire period from Sept. 18, 2015 to May 7, 2016. Compared with the candidates who have dropped out of the race, the three remaining presidential candidates also have the most Twitter followers (Figure \ref{timeline}). This variable is updated every 10 minutes. 

Besides the number of followers, our dataset \textit{US2016} also  contains the detailed follower IDs for Trump, Clinton and Sanders on specific dates, including March 24th, April 17th and May 10th. This information enables us to track the evolution of the election dynamics. We report the summary statistics in Table \ref{sum-id}.

\begin{table}[!h]
\centering
\setlength{\tabcolsep}{8pt}
\renewcommand{\arraystretch}{0.9}
\caption{Number of Followers}
\label{sum-id}
\begin{tabular}{llll}
\hline\hline
                & March 24 & April 17 & May 10  \\\hline
Bernie Sanders  & 1,777,861  & 1,977,982  & 2,134,917 \\
Hillary Clinton & 5,755,618  & 5,905,124  & 6,176,731 \\
Donald Trump    & 7,075,507  & 7,604,915  & 8,020,568 \\\hline
\end{tabular}
\end{table}

Using follower information, we first study among Sanders followers who are following Trump but not Clinton and who are following Clinton but not Trump. We think it is reasonable to assume that if Sanders drops out of the race, Sanders followers who are following Trump but not Clinton will support Trump, and that those who are following Clinton but not Trump will support Clinton. To match the candidates' millions of followers, we first sort their IDs and then use binary search. The entire matching process can be done within a few minutes.\footnote{Codes and data used in this paper are available on the first author's website.}

Furthermore, we collect the profile images based on follower IDs. Our goal is to infer an individual's gender based on the profile image and to test the hypothesis that individuals who follow both Sanders and Trump are more likely to be male than an average Sanders follower.

To process the profile images, we first use OpenCV to identify faces, as the majority of profile images only contain a face.\footnote{http://opencv.org/.} We discard images that do not contain a face and the ones in which OpenCV is not able to detect a face. When multiple faces are available, we choose the largest one. Out of all facial images thus obtained, we select only the large ones. Here we set the threshold to 18kb. This ensures high image quality and also helps remove empty faces. Lastly we resize those images to (28, 28, 3). Eventually, we get 40,088 images sampled from all Sanders followers and 34,921 images from Sanders followers who also follow Trump (Table \ref{image}). % In Table \ref{image}, we report the summary statistics of the images used in classification.

\begin{table}[!h]
\centering
\caption{Number of Profile Images in \textit{US2016}}
\setlength{\tabcolsep}{7.5pt}
\renewcommand{\arraystretch}{0.9}
\label{image}
\begin{tabular}{lll}
\hline\hline
              & \multicolumn{1}{c}{All Sanders} & \multicolumn{1}{c}{Sanders \& Trump} \\\hline
 Number of Images            & \multicolumn{1}{c}{40,088}           & \multicolumn{1}{c}{34,921}     \\
\hline       
\end{tabular}
\end{table}

To classify profile images, we train a convolutional neural network using 42,554 weakly labeled images, with a gender ratio of 1:1. These images come from Trump's and Clinton's current followers. And we infer their labels using the followers' names. For example, David, John, Luke and Michael are male names, and Caroline, Elizabeth, Emily, Isabella and Maria are female names.\footnote{The full list of label names together with the validation data set and the trained model, is available at the first author's website.} For validation, we use a manually labeled data set of 1,965 profile images for gender classification. The validation images come from Twitter as well so that we can avoid the cross-domain problem. Moreover, they do not intersect with the training samples as they come exclusively from individuals who unfollowed Hillary Clinton before March 2016.

\begin{table}[!h]
\centering
\caption{Summary Statistics of CNN Performance}
\label{Performance}
\setlength{\tabcolsep}{6.5pt}
\renewcommand{\arraystretch}{0.9}
\begin{tabular}{lllll}
\hline\hline
Architecture & Precision & Recall & F1    & Accuracy \\
2CONV-1FC    & 91.36     & 90.05  & 90.70 & 90.18   \\
\hline
\end{tabular}
\end{table}

\begin{figure*}[!htbp]
\centering
\includegraphics[height=5cm,width=14cm]{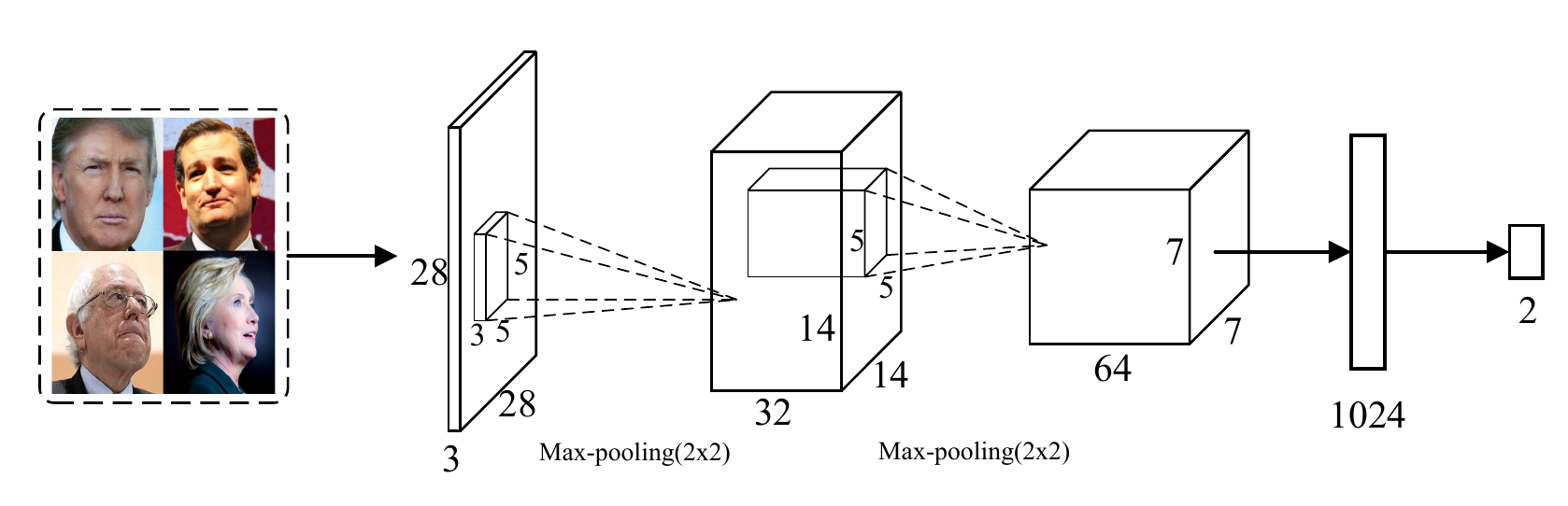}
\caption{The CNN model consists of 2 convolution layers, 2 max-pool layers and a fully connected layer.}
\label{visio}
\end{figure*} 
The architecture of our convolutional neural network is illustrated in Figure \ref{visio}, and the performance of the model is reported in Table \ref{Performance}.

\section{Main Results}
In this section, we answer the question of whether Sanders followers are jumping ship for Trump. Specifically, we examine whether an increasing proportion of Sanders followers are now following Trump and, if so, whether this phenomenon is particularly significant for men.

\subsection{}
In this subsection, we analyze the composition of Sanders followers between March and May. We divide Sanders followers into four groups: (1) only follow Trump and Sanders, (2) only follow Clinton and Sanders, (3) follow Trump, Clinton and Sanders, (4) only follow Sanders. We assume that followers in Group 1 are the most likely to switch to Trump and followers in Group 2 are the least likely. We report our results in Figures 3, 4, 5.

\begin{figure}[!h]
\centering
\includegraphics[height=4cm,width=8.4cm]{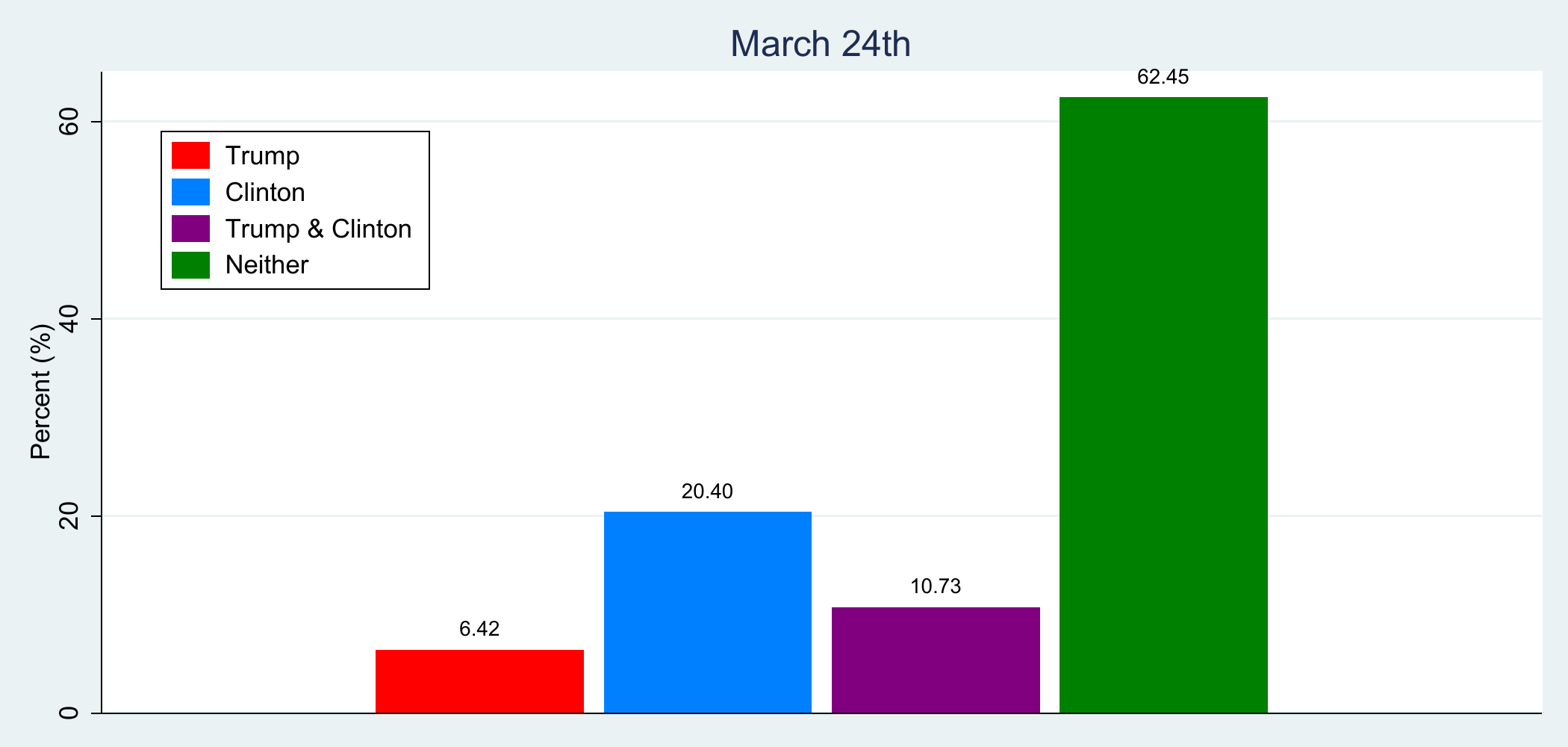}
\caption{Composition of Sanders followers, March.}
\label{march}
\end{figure}

\begin{figure}[!h]
\centering
\includegraphics[height=4cm,width=8.4cm]{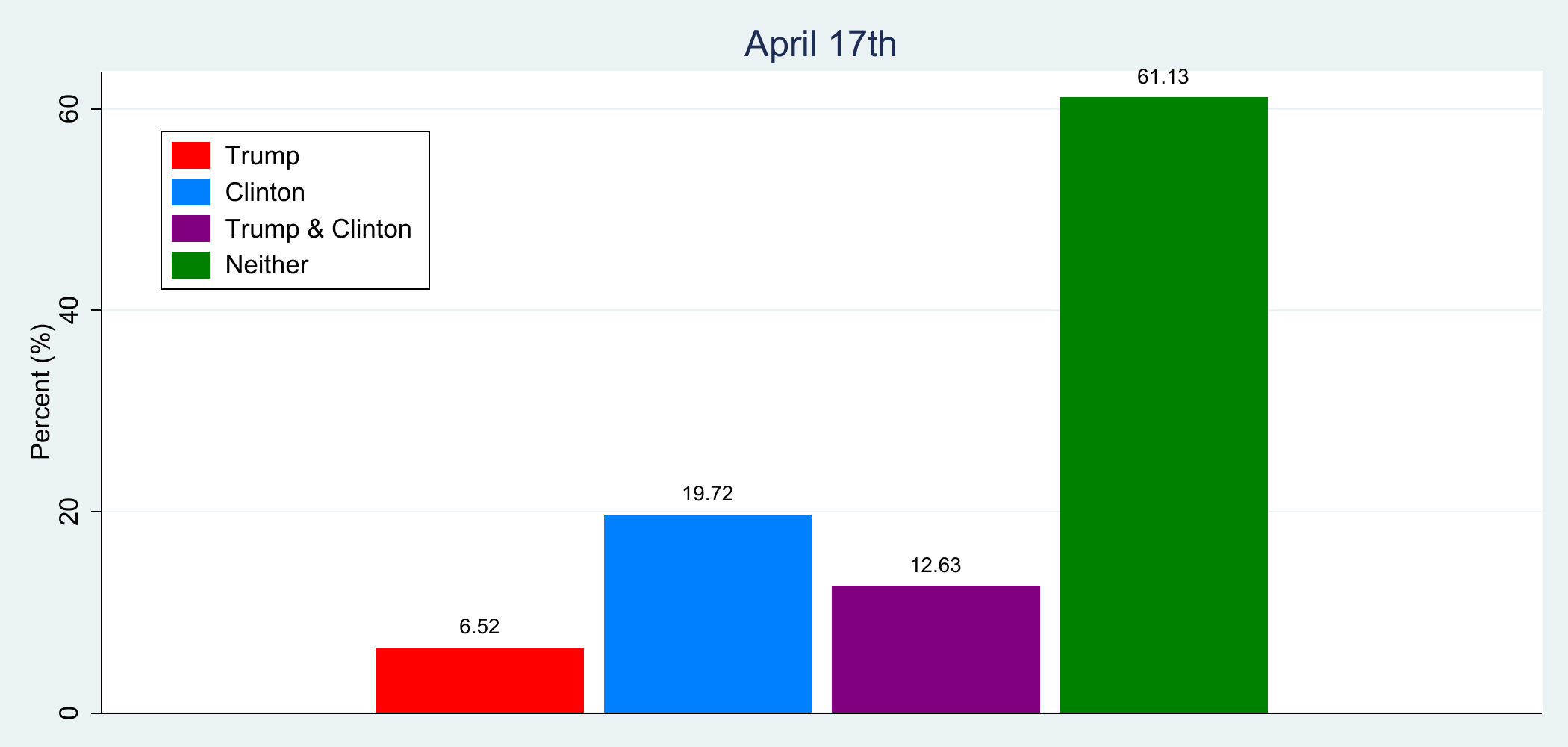}
\caption{Composition of Sanders followers, April.}
\label{woman-follow-clinton}
\end{figure}

\begin{figure}[!htbp]
\centering
\includegraphics[height=4cm,width=8.4cm]{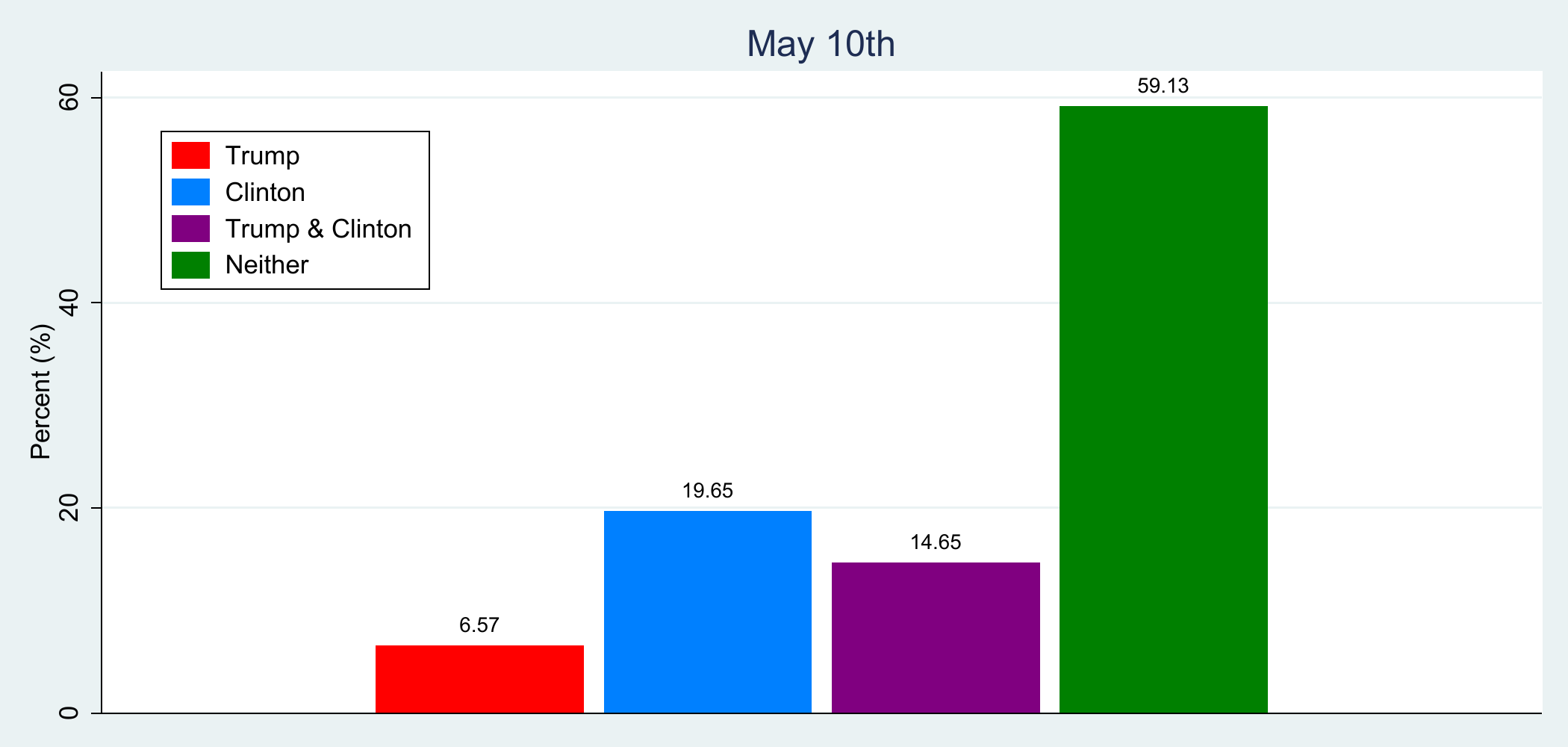}
\caption{Composition of Sanders followers, May.}
\label{woman-follow-clinton}
\end{figure}

\iffalse
\begin{figure}[!htbp]
\centering
\includegraphics[height=20cm,width=8.4cm]{dream-1.pdf}
\caption{Time series Evolution of Sanders followers.}
\label{woman-follow-clinton}
\end{figure}
\fi

The results indicate a decrease of Clinton followers and an increase of Trump followers among Sanders followers between March and May. In Table \ref{stgender}, we use score test to show that the increase of Trump's presence and the drop of Clinton's presence are statistically significant.\footnote{The formula for the score test statistic is: $z=\frac{\hat{p}_1-\hat{p}_2}{\sqrt{\hat{p}(1-\hat{p})(1/n_1+1/n_2)}}$, where $\hat{p}_1=\frac{x}{n_1},\:\hat{p}_2=\frac{y}{n_2},\:p=\frac{x+y}{n_1+n2}.$ With large $n_1$ and $n_2$, z is approximately standard normal.}  

Meanwhile, we also observe that individuals who follow only Sanders, marked by green, make up a smaller share in May than in March and that the share of individuals who follow all the three candidates has increased. Using score test, we are also able to show that these changes are statistically significant. 
\begin{table}[h!]
\centering
\caption{The Composition of Sanders Followers}\label{stgender}
\setlength{\tabcolsep}{3.5pt}
\renewcommand{\arraystretch}{1}
\begin{tabular}{lllll}\hline\hline
\multirow{2}{*}{Null Hypothesis}& \multicolumn{2}{c}{Clinton \& Sanders} & \multicolumn{2}{c}{Trump \& Sanders} \\
\cline{2-3}\cline{4-5}
                                 & z statistic     & \textit{p}  value       & z statistic     & \textit{p}  value \\\hline
Pr$_{March}$=Pr$_{May}$      & -18.47           &  0.00  & 5.99           &  0.00   \\\hline
\end{tabular}
\end{table}

\subsection{}
%and is related to previous studies on the gender effects in Trump's campaign \cite{womancard,votingfeet}
In this subsection, we study whether men are more likely to jump ship for Trump than women. Our investigation is motivated by poll findings that show white male supporters of Bernie Sanders are the most likely to switch to Trump.\footnote{Washington Post, https://www.washingtonpost.com/news/

the-fix/wp/2016/05/24/how-likely-are-bernie-sanders-supporters-to-actually-vote-for-donald-trump-here-are-some-clues.}
 
\begin{figure}[!h]
\centering
\includegraphics[height=4.3cm,width=8.4cm]{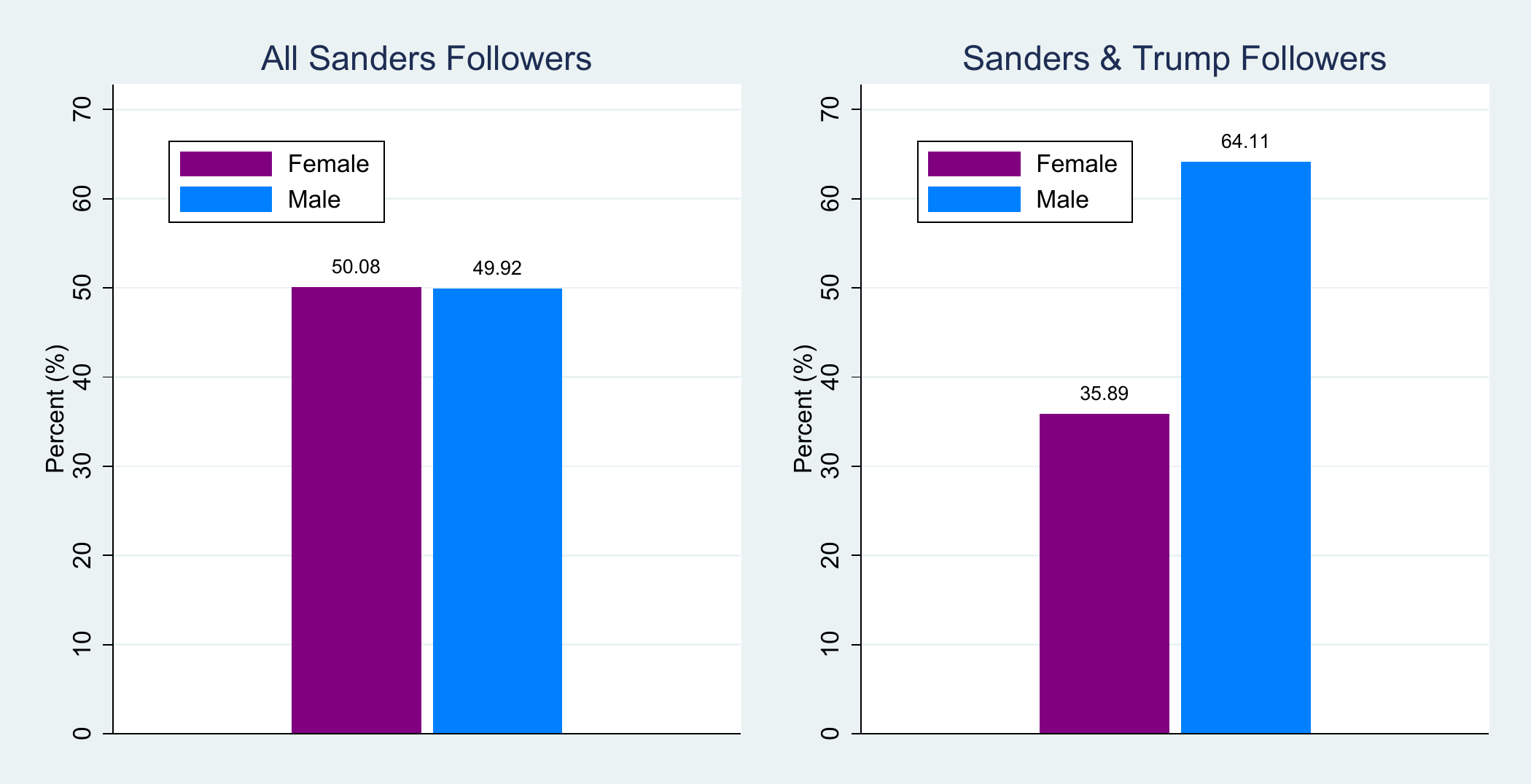}
\caption{Sanders followers who are likely to jump ship for Trump are disproportionately male.}
\label{woman-unfollow-clinton}
\end{figure}

We use data collected on May 10th, when Sanders has 2,134,917 followers, of which 140,185 simultaneously follow Trump but not Clinton. Using the neural network reported in Section 3, we classify the gender of these followers. We report the results in Figure \ref{woman-unfollow-clinton}.

We find that of all Sanders followers 49.92\% are male, but for those who follow both Sanders and Trump (and not Clinton) the percentage is as high as 64.11\%. Using score test (Table \ref{stgender}), we show that among the Sanders followers, those who simultaneously follow Trump but not Clinton are more likely to be male than an average Sanders follower. Our results are consistent with the poll results and lend further support to previous studies that demonstrate the gender effect.

\begin{table}[!h]
\centering
\caption{Score Test on Gender Composition}
\label{stgender}
\setlength{\tabcolsep}{9.2pt}
\renewcommand{\arraystretch}{0.9}
\begin{tabular}{lll}
\hline\hline
\multirow{2}{*}{Null Hopythesis} & \multicolumn{2}{c}{Men} \\
\cline{2-3}
                                 & z statistic     & \textit{p} value      \\\hline
Pr$_{Sanders}$=Pr$_{Sanders\: \&\: Trump}$                            & 39.10           & 0.00   \\\hline
\end{tabular}
\end{table}

\section{Conclusions}
As Hillary Clinton moves closer and closer to clinch the Democratic nomination, the question of whether Sanders supporters would jump ship for Trump becomes decisive. A number of polls suggest that Sanders supporters could end up voting for Trump and that this is particularly so for his male supporters. In this paper, we explored this new dynamic in social media. Building from a unique dataset of the three candidates' Twitter followers, we first analyzed the evolution in the composition of Sanders followers and then using neural network we studied whether there is a gender effect when Sanders supporters consider jumping ship for Trump.

Our study shows that between March and May an increasing proportion of Sanders followers are following Trump (but not Clinton). The proportion of Sanders followers who follow Clinton but not Trump has actually decreased. Equally important, our study suggests that the jumping ship behavior will be affected by the gender effect and that men are more likely to switch to Trump than women. 

%ACKNOWLEDGMENTS are optional
\section{Acknowledgment}
%We gratefully acknowledge support from the University and from our corporate sponsors.

We gratefully acknowledge support from the University and from our corporate sponsors Xerox and Yahoo.
\bibliographystyle{acm}
\bibliography{yu} 

\begin{thebibliography}{10}

\bibitem{genderAffinityEffect}
{\sc Dolan, K.}
\newblock {I}s {T}here a ``{G}ender {A}ffinity {E}ffect" in {A}merican
  {P}olitics? {I}nformation, {A}ffect, and {C}andidate {S}ex in {U}.{S}.
  {H}ouse {E}lections.
\newblock {\em Political Research Quarterly\/} (2008).

\bibitem{facerace}
{\sc Fu, S., He, H., and Hou, Z.-G.}
\newblock {L}earning {R}ace from {F}ace: {A} {S}urvey.
\newblock In {\em Pattern Analysis and Machine Intelligence, IEEE Transactions
  on\/} (2014), vol.~36.

\bibitem{ginosar}
{\sc Ginosar, S., Rakelly, K., Sachs, S., Yin, B., and Efros, A.~A.}
\newblock {A} {C}entury of {P}ortraits: {A} {V}isual {H}istorical {R}ecord of
  {A}merican {H}igh {S}chool {Y}earbooks.
\newblock In {\em ICCV 2015 Extreme Imaging Workshop Proceedings\/} (2015).

\bibitem{sexAndGOP}
{\sc King, D.~C., and Matland, R.~E.}
\newblock {S}ex and the {G}rand {O}ld {P}arty: {A}n {E}xperimental
  {I}nvestigation of the {E}ffect of {C}andidate {S}ex on {S}upport for a
  {R}epublican {C}andidate.
\newblock {\em American Politics Research\/} (2003).

\bibitem{israel}
{\sc Levi, G., and Hassner, T.}
\newblock {A}ge and {G}ender {C}lassification using {D}eep {C}onvolutional
  {N}eural {N}etworks.
\newblock In {\em {P}roceedings of the {IEEE} {C}onference on {C}omputer
  {V}ision and {P}attern {R}ecognition\/} (2015), pp.~34--42.

\bibitem{facebookCongress}
{\sc MacWilliams, M.~C.}
\newblock {F}orecasting {C}ongressional {E}lections {U}sing {F}acebook {D}ata.
\newblock {\em PS: Political Science \& Politics 48}, 04 (October 2015).

\bibitem{tweets2polls}
{\sc O'Connor, B., Balasubramanyan, R., Routledge, B.~R., and Smith, N.~A.}
\newblock {F}rom {T}weets to {P}olls: {L}inking {T}ext {S}entiment to {P}ublic
  {O}pinion {T}ime {S}eries.
\newblock In {\em Proceedings of the Fourth International AAAI Conference on
  Weblogs and Social Media\/} (2010).

\bibitem{womancard}
{\sc Wang, Y., Feng, Y., Li, Y., Zhang, X., Niemi, R., and Luo, J.}
\newblock {P}ricing the {W}oman {C}ard: {G}ender {P}olitics between {H}illary
  {C}linton and {D}onald {T}rump.
\newblock In {\em {I}n ar{X}iv preprint:1605.05401\/} (2016).
\newblock In arXiv prepring:1605.05401 (2016).

\bibitem{trumpists}
{\sc Wang, Y., Li, Y., and Luo, J.}
\newblock {D}eciphering the 2016 {U}.{S}. {P}residential {C}ampaign in the
  {T}witter {S}phere: {A} {C}omparison of the {T}rumpists and {C}lintonists.
\newblock In {\em {T}enth {I}nternational {AAAI} {C}onference on {W}eb and
  {S}ocial {M}edia\/} (2016).

\bibitem{votingfeet}
{\sc Wang, Y., Li, Y., You, Q., Zhang, X., Niemi, R., and Luo, J.}
\newblock {V}oting with {F}eet: {W}ho are {L}eaving {H}illary {C}linton and
  {D}onald {T}rump?
\newblock In {\em ar{X}iv preprint :1604.07103\/} (2016).

\bibitem{neco}
{\sc Wang, Y., Luo, J., Niemi, R., and Li, Y.}
\newblock {T}o {F}ollow or {N}ot to {F}ollow: {A}nalyzing the {G}rowth
  {P}atterns of the {T}rumpists on {T}witter.
\newblock In {\em {W}orkshop {P}roceedings of the 10th {I}nternational {AAAI}
  {C}onference on {W}eb and {S}ocial {M}edia\/} (2016).

\bibitem{trumponfire}
{\sc Wang, Y., Luo, J., Niemi, R., Li, Y., and Hu, T.}
\newblock {C}atching {F}ire via `{L}ikes': {I}nferring {T}opic {P}references of
  {T}rump {F}ollowers on {T}witter.
\newblock In {\em {T}enth {I}nternational {AAAI} {C}onference on {W}eb and
  {S}ocial {M}edia\/} (2016).

\end{thebibliography}

\end{document}